\documentclass[11pt]{article}

\usepackage[a4paper, margin=1in]{geometry}
\usepackage{amsmath, amssymb, amsthm}
\usepackage{stmaryrd}
\usepackage{graphicx}
\usepackage{hyperref}
\usepackage{natbib}
\usepackage{setspace}

\newtheorem{lemma}{Lemma}
\newtheorem{theorem}{Theorem}

\newtheorem{corollary}{Corollary}

\title{\textbf{Depth-13 Sorting Networks for 28 Channels}}
\author{
  Chengu Wang \\
  \small{\href{mailto:wangchengu@gmail.com}{wangchengu@gmail.com}} \\
}
\date{}

\begin{document}

\maketitle

\begin{abstract}
  We establish new depth upper bounds for sorting networks on 27 and 28 channels, improving the previous best bound of 14 to 13. Our 28-channel network is constructed with reflectional symmetry by combining high-quality prefixes of 16- and 12-channel networks, extending them greedily one comparator at a time, and using a SAT solver to complete the remaining layers.
\end{abstract}

\section{Introduction}

A comparator network is a circuit in which each gate is a comparator that takes two inputs and produces the minimum value (on top) and maximum value (on bottom) as outputs, illustrated as a vertical line in network diagrams. As with any circuit, we are concerned with its size and depth. The size of a comparator network is the number of comparators it contains. The depth of a comparator network is the maximum number of comparators along any path from input to output. A sorting network is a comparator network that sorts its inputs into non-decreasing order.

\begin{figure}[h]
  \centering
  \includegraphics[width=0.2 \textwidth]{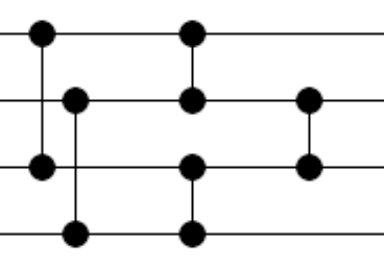}
  \caption{A sorting network for 4 channels of depth 3.}
  \label{fig:sorting-network-4}
\end{figure}

For example, Figure~\ref{fig:sorting-network-4} shows a sorting network with 4 channels of depth 3.

Finding optimal sorting networks is crucial for practical applications. Hardware implementations of sorting networks appear in GPU sorting, FPGA-based systems, and cryptographic protocols. Software-based sorting networks serve as building blocks for oblivious sorting algorithms used in secure computation and differential privacy systems.

Batcher developed two algorithms---odd-even mergesort and bitonic mergesort---that construct networks with depth $O((\log n)^2)$ and size $O(n (\log n)^2)$ \cite{batcher1968sorting,knuth1997artv3}. A significant theoretical breakthrough was later achieved by Ajtai, Komlós, and Szemerédi with their AKS network construction, which attains depth $O(\log n)$ and size $O(n \log n)$ \cite{ajtai19830}. While these constructions are asymptotically optimal, the large constants hidden by the Big-O notation render them impractical for real-world applications. Furthermore, the theoretical understanding gained from studying small optimal networks can inform the design of more efficient asymptotic constructions.

For this reason, our work focuses on finding optimal-depth networks for specific small values of $n$. Table \ref{tab:bounds} shows the best-known lower and upper bounds on the depth of sorting networks for small input sizes $n$ prior to our work.

\begin{table}[h]
  \centering
  \caption{Known depth lower and upper bounds for $n=1$ to $32$}
  \label{tab:bounds}
  \begin{tabular}{|l|l|l|}
    \hline
    $n$ & depth LB                      & depth UB                            \\
    \hline
    1   & 0 \cite{knuth1997artv3}       & 0 \cite{knuth1997artv3}             \\
    \hline
    2   & 1 \cite{knuth1997artv3}       & 1 \cite{knuth1997artv3}             \\
    \hline
    3   & 3 \cite{knuth1997artv3}       & 3 \cite{knuth1997artv3}             \\
    \hline
    4   & 3 \cite{knuth1997artv3}       & 3 \cite{knuth1997artv3}             \\
    \hline
    5   & 5 \cite{knuth1997artv3}       & 5 Shapiro \cite{knuth1997artv3}     \\
    \hline
    6   & 5 \cite{knuth1997artv3}       & 5 Shapiro \cite{knuth1997artv3}     \\
    \hline
    7   & 6 \cite{knuth1997artv3}       & 6 \cite{knuth1997artv3}             \\
    \hline
    8   & 6 \cite{knuth1997artv3}       & 6 \cite{knuth1997artv3}             \\
    \hline
    9   & 7 \cite{parberry1989computer} & 7 Shapiro \cite{knuth1997artv3}     \\
    \hline
    10  & 7 \cite{parberry1989computer} & 7 Van Voorhis \cite{knuth1997artv3} \\
    \hline
    11  & 8 \cite{bundala2014optimal}   & 8 Shapiro \cite{knuth1997artv3}     \\
    \hline
    12  & 8 \cite{bundala2014optimal}   & 8 Shapiro \cite{knuth1997artv3}     \\
    \hline
    13  & 9 \cite{bundala2014optimal}   & 9 Van Voorhis \cite{knuth1997artv3} \\
    \hline
    14  & 9 \cite{bundala2014optimal}   & 9 Van Voorhis \cite{knuth1997artv3} \\
    \hline
    15  & 9 \cite{bundala2014optimal}   & 9 Van Voorhis \cite{knuth1997artv3} \\
    \hline
    16  & 9 \cite{bundala2014optimal}   & 9 Van Voorhis \cite{knuth1997artv3} \\
    \hline
  \end{tabular}
  \hspace{0.5cm}
  \begin{tabular}{|l|l|l|}
    \hline
    $n$            & depth LB                    & depth UB                    \\
    \hline
    17             & 10 \cite{codish2019sorting} & 10 \cite{ehlers2014faster}  \\
    \hline
    18             & 10 \cite{codish2019sorting} & 11 \cite{al2009finding}     \\
    \hline
    19             & 10 \cite{codish2019sorting} & 11 \cite{ehlers2014faster}  \\
    \hline
    20             & 10 \cite{codish2019sorting} & 11 \cite{ehlers2014faster}  \\
    \hline
    21             & 10 \cite{codish2019sorting} & 12 \cite{al2009finding}     \\
    \hline
    22             & 10 \cite{codish2019sorting} & 12 \cite{al2009finding}     \\
    \hline
    23             & 10 \cite{codish2019sorting} & 12 \cite{ehlers2017merging} \\
    \hline
    24             & 10 \cite{codish2019sorting} & 12 \cite{ehlers2017merging} \\
    \hline
    25             & 10 \cite{codish2019sorting} & 13 \cite{sorterhunter}      \\
    \hline
    26             & 10 \cite{codish2019sorting} & 13 \cite{sorterhunter}      \\
    \hline
    27$^{\dagger}$ & 10 \cite{codish2019sorting} & 14 Odd-Even Merge           \\
    \hline
    28$^{\dagger}$ & 10 \cite{codish2019sorting} & 14 Odd-Even Merge           \\
    \hline
    29             & 10 \cite{codish2019sorting} & 14 Odd-Even Merge           \\
    \hline
    30             & 10 \cite{codish2019sorting} & 14 Odd-Even Merge           \\
    \hline
    31             & 10 \cite{codish2019sorting} & 14 Odd-Even Merge           \\
    \hline
    32             & 10 \cite{codish2019sorting} & 14 Odd-Even Merge           \\
    \hline
  \end{tabular}

  \vspace{5pt}
  \footnotesize{$\dagger$ indicates the number of channels this paper is interested in.}
\end{table}

The lower bounds for $n \leq 17$ are known to be tight. Look at the column ``depth UB" in the right table. 10 appears once, 11 appears three times, 12 appears four times, 13 appears twice, and 14 appears six times. Numbers 10 and 13 are fewer than normal, and number 14 is more than normal. So, it suggests that depth upper bounds for $n=18, 27$ and $28$ might not be tight. Unfortunately, we were not able to improve the depth upper bound for 18 channels in this paper. This paper is for 27 and 28 channels.

A sorting network is reflection-symmetric if it is invariant under channel reflection (vertical reflection in the diagram). For example, Figure~\ref{fig:sorting-network-4} is reflection-symmetric. Reflection-symmetric sorting networks are so powerful. For all the best-known sorting networks with minimum depths for even $n \leq 32$, there exists a reflection-symmetric sorting network with the same depth. The reflection-symmetry can be used to reduce the search space when improving the depth upper bound. SorterHunter~\cite{sorterhunter} uses the reflection-symmetry and found many best-known depth upper bounds of sorting networks. Codish, Cruz-Filipe, Schneider-Kamp~\cite{codish2014quest}, Bundala, and Závodnỳ~\cite{bundala2014optimal} also use it to reduce the search space of the prefixes.

Ehlers~\cite{ehlers2017merging} improved the depth upper bound of the 24-channel networks to 12 layers. It stacked the prefixes of two 12-channel networks, and solved the remaining layers by a SAT solver.

Our construction of n=28 and d=13 is based on \cite{ehlers2017merging}. As we have more channels and one more layer, we use two new techniques to reduce the search space: assume the network is reflection-symmetric, and greedily extend the 24-channel prefix by one layer before solving the SAT problem. The full construction is found within 20 minutes on a common desktop computer.

\section{Preliminaries and Redundant Prefixes}
The search space for sorting networks grows exponentially with the number of channels, making exhaustive search computationally prohibitive for larger values of $n$. Consequently, researchers have developed various techniques to prune the search space and accelerate the discovery of optimal networks.

\begin{lemma}[Zero-One Principle]\label{lemma:zero-one-principle}
  If a comparator network on $n$ channels sorts all $2^n$ Boolean sequences into non-decreasing order, then it correctly sorts all sequences of arbitrary (comparable) values. \cite{knuth1997artv3}
\end{lemma}

The zero-one principle is a powerful result that allows us to verify the correctness of a sorting network by testing only the $2^n$ binary input sequences rather than $n!$ cases.

For comparator networks $C$ and $D$ with the same number of channels, denote $C\fatsemi D$ the composition of $C$ and $D$ by putting $D$ after $C$, i.e. for input $x$, $(C\fatsemi D)(x) = D(C(x))$.

For a comparator network $C$, denote $\operatorname{output}(C)=\{ C(x) | x \in \{0,1\}^n \}$ the set of all possible outputs of $C$ on Boolean inputs. The following lemma states that
it suﬃces to consider prefixes P with minimal $\operatorname{output}(P)$.

\begin{lemma}
  Let $P \fatsemi S$ be a sorting network, and $P'$ be a comparator network. If $\operatorname{output}(P') \subseteq \operatorname{output}(P)$, then $P' \fatsemi S$ is a sorting network. \cite{bundala2014optimal}
\end{lemma}

To further reduce the search space, we use some symmetries: permutation and negation.

\begin{lemma}
  Let $P \fatsemi S$ be a sorting network on $n$ channels, and $P'$ be a comparator network on $n$ channels. If there exists a permutation $\sigma \in \mathbb{S}_n$ such that $\sigma (\operatorname{output}(P')) \subseteq \operatorname{output}(P)$, then there exists a comparator network $S'$ with the same depth as $S$ such that $P' \fatsemi S'$ is a sorting network. \cite{bundala2014optimal}
\end{lemma}

We denote the bitwise-not operator by $\neg$. For example, $\neg 001_2 = 110_2$. When it is applied to a set of outputs, it flips each output, e.g. $\neg \{000_2,001_2,010_2\} = \{111_2,110_2,101_2\} = \{101_2,110_2,111_2\}$. Also, For an integer $n$, we denote $[n]$ the set $\{0, 1, \ldots, n-1\}$, and $\rho_n$ the reflection permutation on $[n]$, i.e. $\rho_n(i) = n-1-i$.

\begin{lemma}
  If a comparator network $S$ sorts input $x\in \{0,1\}^n$, then $S$ sorts $\rho_n(\neg x)$.
\end{lemma}
Bundala and Závodnỳ~\cite{bundala2014optimal} use this property to reduce the search space of the prefixes. We are giving a short proof here for completeness.
\begin{proof}
  Denote $\neg S$ the network that changes all the comparators in $S$ to max-min-comparators, which output max on top and min on bottom. Assume $S$ sorts $x$. We flip every bit in the network, so $\neg S$ sorts $\neg x$ into the reverse order. Then, we vertically reflect the network along with the input $x$. Therefore, $S$ sorts $\rho_n(\neg x)$.
\end{proof}

Putting them together, we get the following corollary that allows us to remove the redundant prefixes in the search space.

\begin{corollary}\label{corollary:redundant-prefixes}
  Let $P \fatsemi S$ be a sorting network on $n$ channels, and $P'$ be a comparator network on $n$ channels. If there exists a permutation $\sigma \in \mathbb{S}_n$ such that $\sigma (\operatorname{output}(P')) \subseteq \operatorname{output}(P)$ or $\sigma (\neg \operatorname{output}(P')) \subseteq \operatorname{output}(P)$, then there exists a comparator network $S'$ with the same depth as $S$ such that $P' \fatsemi S'$ is a sorting network. \cite{bundala2014optimal}
\end{corollary}

The lemma states that if the outputs of $P'$ (or their complements) are contained within the outputs of $P$ up to permutation, then we can replace $P$ with $P'$ in the construction. This allows us to prune the search space by removing redundant prefixes.

In practice, it is not trivial to check if a network output set $A$ is isomorphic to the subset of another one $B$. We use two techniques to speed up the check, one for positive cases, and one for negative cases.

Consider $A$ as an $m_A \times n$ Boolean matrix, and $A_{i,j}\in \{0,1\}$ is the output on $j$-th channel of the $i$-th output. Similarly, we consider $B$ as an $m_B \times n$ Boolean matrix. We sort the $n$ columns of $A$ and $B$ by the column sum, respectively. If multiple columns have the same sum, we permute them randomly. If the rows of $B$ covers the rows of $A$, then $A$ is isomorphic to a subset of $B$. We repeat this random process a couple of times. In practice, it detects more than $90\%$ of the redundant outputs efficiently.

To quickly filter negative cases, we compute the column sums and row sums of matrix $A$ and $B$. Denote the row sum sequence of $A$ by $\mathrm{RowSumA}_i$ for $i \in [m_A]$, and the column sum sequence of $A$ by $\mathrm{ColSumA}_j$ for $j \in [n]$. Similarly, we compute $\mathrm{RowSumB}$ and $\mathrm{ColSumB}$. We sort $\mathrm{RowSumA}$, $\mathrm{RowSumB}$, $\mathrm{ColSumA}$, and $\mathrm{ColSumB}$, respectively. If $A$ is isomorphic to a subset of $B$, then $\mathrm{ColSumA}_j \leq \mathrm{ColSumB}_j$ for all $j \in [n]$, and $\mathrm{RowSumA}$ is a subsequence of $\mathrm{RowSumB}$. This property is commonly used in bipartite graph isomorphism and embedding detections and sorting networks \cite{codish2016sorting}.

This paper focuses on reflection-symmetric sorting networks on even channels. We say a permutation $\sigma \in \mathbb{S}_n$ is reflection-symmetric if it is commutative with $\rho_n$, i.e. $\rho_n \circ \sigma = \sigma \circ \rho_n$. All these reflection-symmetric permutations form the centralizer of $\rho_n$ in $\mathbb{S}_n$, denoted by $\mathcal{C}(\rho_n)$. It is isomorphic to the wreath product of $\mathbb{C}_2$ and $\mathbb{S}_{n/2}$. In other words, we can permute the $n/2$ pairs (the $\mathbb{S}_{n/2}$ part), and independently flip each pair (the $\mathbb{C}_2$ part).

By applying the reflection-symmetry on Corollary~\ref{corollary:redundant-prefixes}, we obtain:
\begin{corollary}\label{corollary:redundant-prefixes-reflection}
  Let $P \fatsemi S$ be a reflection-symmetric sorting network on $n$ channels, and let $P'$ be a reflection-symmetric comparator network on $n$ channels. If there exists a reflection-symmetric permutation $\sigma \in \mathcal{C}(\rho_n)$ such that $\sigma (\operatorname{output}(P')) \subseteq \operatorname{output}(P)$ or $\sigma (\neg \operatorname{output}(P')) \subseteq \operatorname{output}(P)$, then there exists a reflection-symmetric comparator network $S'$ with the same depth as $S$ such that $P' \fatsemi S'$ is a sorting network.
\end{corollary}

We will use this corollary to eliminate redundant prefixes in Section~\ref{sec:n28d13}. Specifically, if both $P$ and $P'$ are in our search space, we can remove $P$ and keep only $P'$. The two speedup techniques described above also apply to the reflection-symmetric case by maintaining reflection symmetry at each step.

\section{Construction of Depth-13 Sorting Network for 28 Channels}\label{sec:n28d13}

\begin{figure}[h]
  \centering
  \includegraphics[width=\textwidth]{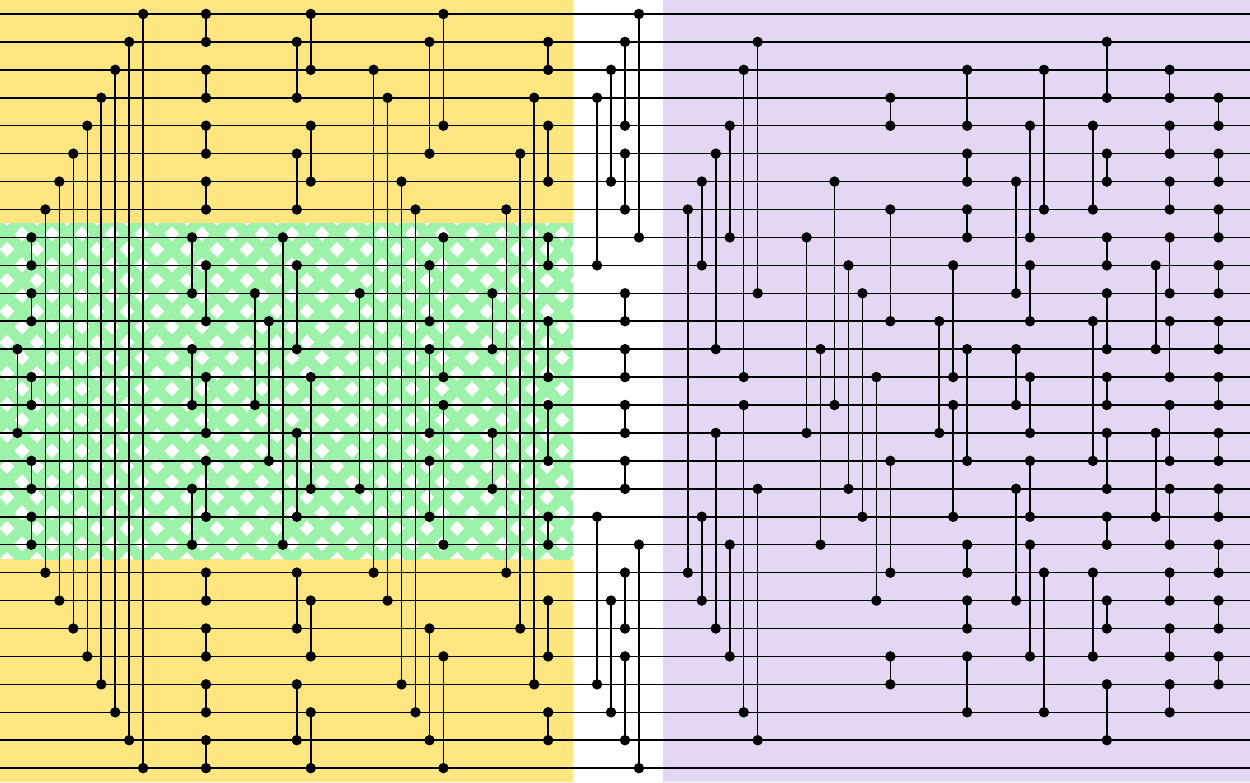}
  \caption{A sorting network for 28 channels with 13 layers}
  \label{fig:n28d13}
\end{figure}

We found a reflection-symmetric sorting network for 28 channels with 13 layers, shown in Figure~\ref{fig:n28d13} and listed below:

\begin{scriptsize}\begin{verbatim}
[(0,27),(1,26),(2,25),(3,24),(4,23),(5,22),(6,21),(7,20),(8,9),(10,11),(12,15),(13,14),(16,17),(18,19)]
[(0,1),(2,3),(4,5),(6,7),(8,10),(9,11),(12,14),(13,15),(16,18),(17,19),(20,21),(22,23),(24,25),(26,27)]
[(0,2),(1,3),(4,6),(5,7),(8,19),(9,12),(10,14),(11,16),(13,17),(15,18),(20,22),(21,23),(24,26),(25,27)]
[(0,4),(1,5),(2,20),(3,21),(6,24),(7,25),(8,13),(9,11),(10,17),(12,15),(14,19),(16,18),(22,26),(23,27)]
[(1,2),(3,24),(4,6),(5,22),(7,20),(8,9),(10,12),(11,13),(14,16),(15,17),(18,19),(21,23),(25,26)]
[(0,8),(1,4),(2,6),(3,9),(5,7),(10,11),(12,13),(14,15),(16,17),(18,24),(19,27),(20,22),(21,25),(23,26)]
[(1,10),(2,13),(4,8),(5,12),(6,9),(7,20),(14,25),(15,22),(17,26),(18,21),(19,23)]
[(3,4),(6,14),(7,11),(8,15),(9,17),(10,18),(12,19),(13,21),(16,20),(23,24)]
[(2,4),(5,6),(7,8),(9,13),(11,15),(12,16),(14,18),(19,20),(21,22),(23,25)]
[(2,7),(4,8),(6,10),(9,11),(12,14),(13,15),(16,18),(17,21),(19,23),(20,25)]
[(1,3),(4,7),(5,6),(8,9),(10,12),(11,16),(13,14),(15,17),(18,19),(20,23),(21,22),(24,26)]
[(2,3),(4,5),(6,7),(8,10),(9,12),(11,13),(14,16),(15,18),(17,19),(20,21),(22,23),(24,25)]
[(3,4),(5,6),(7,8),(9,10),(11,12),(13,14),(15,16),(17,18),(19,20),(21,22),(23,24)]  
\end{verbatim}\end{scriptsize}

Our construction strategy consists of three phases: we stack a high-quality 16-channel 5-layer prefix (shown in green in the figure) with a high-quality 12-channel 5-layer prefix (shown in yellow), extend to 6 layers by greedily adding one comparator at a time while maintaining a small set of promising candidates (shown in white), and finally use an SAT solver to figure out layers 7 through 13 (shown in purple).

Merging is most efficient when the two parts have approximately equal numbers of channels. For $n=28$, both $14+14$ and $16+12$ are promising decompositions. Networks are typically more efficient when the number of channels is a power of 2 or a multiple thereof. We chose the $16+12$ decomposition based on these considerations.

\subsection{High-Quality 16-Channel and 12-Channel 5-Layer Prefixes}

We enumerate all the non-redundant reflection-symmetric 12-channel 5-layer prefixes by the generate-and-prune approach \cite{codish2016sorting} one layer at a time.

Starting from the set containing only the empty network, we extend the prefix set by one layer by enumerating all possible reflection-symmetric layers, then prune the resulting set using Corollary~\ref{corollary:redundant-prefixes-reflection}. This process yields 2164 non-redundant 12-channel 5-layer prefixes.

\begin{center}
  \begin{tabular}{|c|c|c|c|c|c|c|}
    \hline
    prefix depth              & 0 & 1 & 2  & 3    & 4     & 5    \\
    \hline
    \# non-redundant prefixes & 1 & 4 & 41 & 1502 & 11753 & 2164 \\
    \hline
  \end{tabular}
\end{center}

We consider a prefix $P$ high-quality if the size of $\operatorname{output}(P)$ is small. Among the 2164 non-redundant prefixes, we select the best 4 prefixes, which have output sizes of 34, 34, 35, and 35. All other prefixes have output sizes larger than 35.

\begin{figure}[h]
  \centering
  \includegraphics[width=0.4 \textwidth]{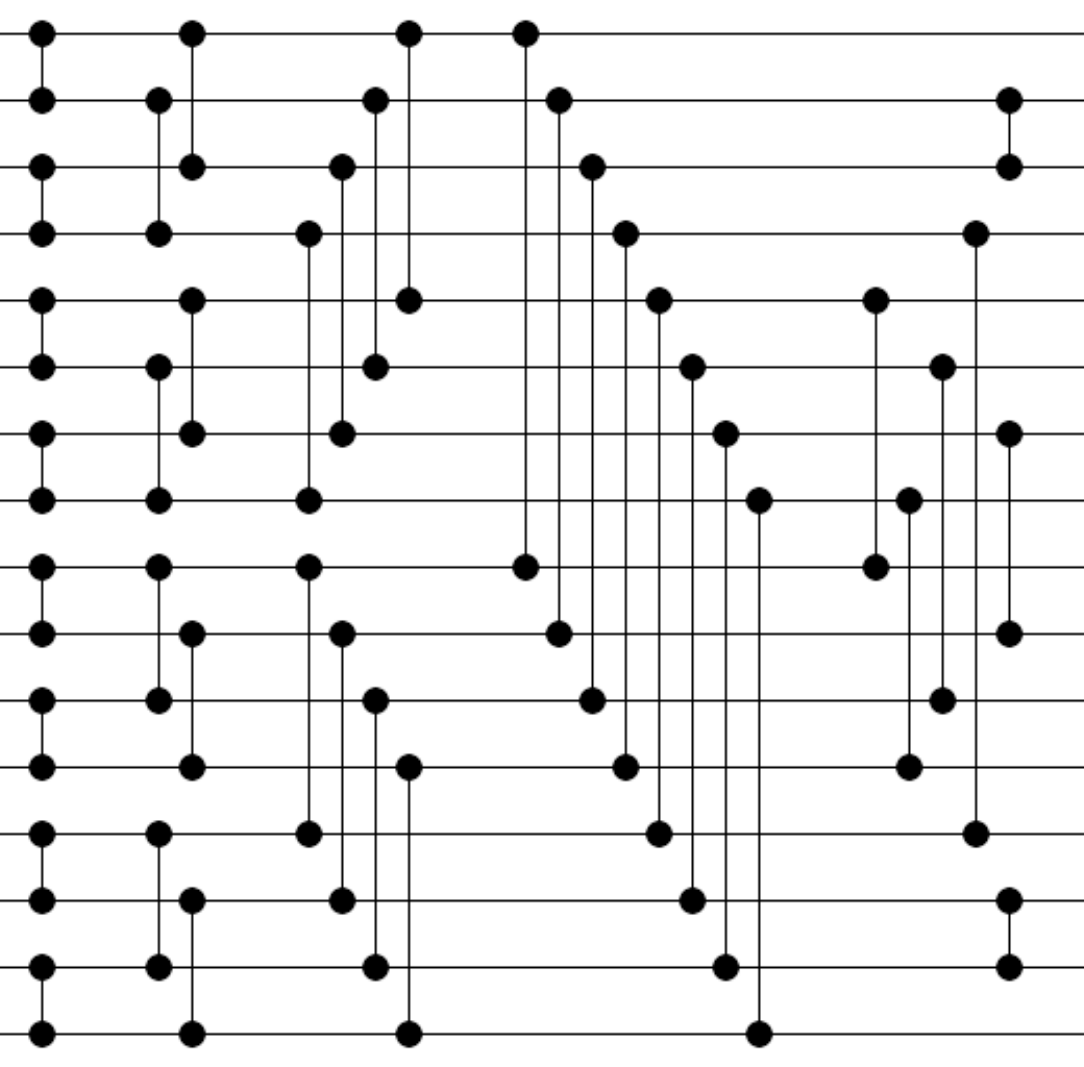}
  \caption{The first 5 layers of Van Voorhis's sorting network for 16 channels}
  \label{fig:n16d5vv}
\end{figure}

For $n=16$, enumerating all non-redundant 5-layer prefixes is computationally challenging. However, there exists an excellent choice: the first 5 layers of Van Voorhis's sorting network, shown in Figure~\ref{fig:n16d5vv}. The first four layers construct a 4-dimensional hypercube structure on the 16 channels. The 5th layer compares channels with the same Hamming weight symmetrically. This structure becomes clearer when we write the 5th layer in binary format: (0001,0010),(0100,1000); (0011,1100),(0101,1010),(0110,1001); (0111,1011),(1101,1110). This design preserves as much information as possible by comparing corresponding keys \cite{al2009finding}. We found two other output sets isomorphic to this one but not isomorphic under reflection-symmetric permutations in $\mathcal{C}(\rho_{16})$.

With the best 4 prefixes for 12 channels and the best 2 prefixes for 16 channels, we combine them to obtain $4 \times 2 = 8$ prefixes for 28 channels with 5 layers.

\subsection{Extending to 6 Layers}

With 5 layers in hand, solving layers 6 through 13 directly with an SAT solver is too slow, even when restricting to reflection-symmetric solutions. However, solving layers 7 through 13 is feasible. Therefore, we extend the prefix by one additional layer. Rather than enumerating all possible symmetric layers on 28 channels---which would be prohibitively expensive---we extend the network one comparator at a time.

We begin with the 8 prefixes $\mathcal{P}_0$. At each step, we add one comparator along with its reflection (if not self-symmetric): $\mathcal{Q}_{i+1} = \{P \fatsemi \{c, \rho_{28}(c)\} \mid P \in \mathcal{P}_i$ and $c$ is a comparator connecting two unused channels in layer $6\}$. We then prune $\mathcal{P}_{i} \cup \mathcal{Q}_{i+1}$ using Corollary~\ref{corollary:redundant-prefixes-reflection} and retain only the best 64 prefixes $\mathcal{P}_{i+1}$---those with the smallest output sizes. We repeat this process until obtaining $\mathcal{P}_{14}$, the set of 64 prefixes with 6 layers.

\subsection{SAT Solving the Remaining Layers}

We now solve layers 7 through 13 using an SAT solver. We employ many optimization techniques from \cite{codish2019sorting} to accelerate the SAT solving process, including improved SAT encodings with oneUp and oneDown techniques, constraints on the last two layers, and channel permutation to minimize the window size.

In practice, it suffices to solve for just the best 8 prefixes in $\mathcal{P}_{14}$. All 8 prefixes yield solutions. We then remove unused comparators---those whose first input is always less than or equal to the second input. Figure~\ref{fig:n28d13} shows one such solution after removing unused comparators and permuting the input channels.

\begin{theorem}
  There exists a reflection-symmetric sorting network for 28 channels with 13 layers.
\end{theorem}

\begin{corollary}
  There exists a sorting network for 27 channels with 13 layers.
\end{corollary}

\subsection{Implementation and Running Time}

We ran our program on a Mac mini M2 with 16GB RAM. The entire computation takes less than 20 minutes to find the solution shown in Figure~\ref{fig:n28d13}. Specifically, finding the best 4 prefixes for 12 channels with 5 layers takes 7 minutes, while finding the best 2 prefixes for 16 channels with 5 layers takes 1 minute. We use MiniSat to solve the 8 SAT instances in parallel. It takes half a minute to find one solution, and 5 minutes to solve all 8 problems. All other steps complete very quickly.

The complete program is available at \url{https://github.com/wcgbg/sorting-network-n28d13}.

\section{Conclusion}

In this paper, we improved the depth upper bound of sorting networks for 27 and 28 channels from 14 to 13. Our approach employs two key techniques to reduce the search space: (1) restricting the search to reflection-symmetric networks, and (2) extending the 6th layer incrementally by adding one comparator at a time while maintaining a small set of high-quality candidates.

Future work could explore more sophisticated prefix selection strategies, such as using reinforcement learning to predict which prefixes are most likely to yield solutions. In this paper, we used minimum output size as the objective function in the greedy extension step. A machine learning model could potentially identify better objective functions that lead to more efficient discovery of optimal networks.

% --- References ---
\bibliographystyle{alpha}
\bibliography{references}

\end{document}